# Реализация виртуальных структур на едином поле файловой системы

к.т.н. М.А. Малков

## Введение

В статье предлагается новый подход к построению иерархической структуры файловой системы. В первой части статьи сделан обзор построения структур хранения файлов в современных операционных системах. Во второй части описывается новый подход к построению структур файловой системы (виртуальные структуры). Такой подход позволяет строить различные системообразующие структуры из одного и того же множества файлов по разным признакам в виде множественных деревьев.

### 1. Структура хранения файлов в современных ОС

Доступ и хранение обрабатываемой информации на ЭВМ реализуется через механизм файловой системы[1]. Для этого операционная система предоставляет возможность построения логической структуры файловой системы, которая отображается на структуру физических носителей для долговременного хранения. Различные операционные системы имеют разную структуру отображения и доступа на физическом уровне, но для пользователя файловая система – это некоторая абстрактная структура, представляемая в виде дерева, узлами которого могут быть файлы либо каталоги. Каталог может содержать внутри себя файлы или (и) другие каталоги (ссылки на файлы или другие каталоги), которые в свою очередь так же могут содержать в себе другие файлы или каталоги и т.д. Таким образом, выстраивается древовидная структура, представленная на рис. 1.

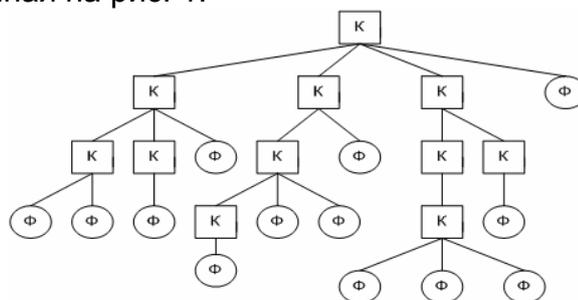

*Рис. 1 – Древовидная структура файловой системы.*

Файл при такой организации структуры имеет, так называемое, короткое имя, которое идентифицирует его внутри каталога, где он расположен. Кроме того, файл имеет полное имя, представляющее собой цепочку коротких имён всех каталогов, через которые проходит путь от корня до самого файла. Таким образом, между полным именем файла и самим файлом имеется взаимно однозначное соответствие. Файл – это элемент древовидной структуры, местоположение которого определяется по полному имени в этой структуре. Такая древовидная структура позволяет:
- использовать одинаковые короткие имена файлов, если они расположены в разных каталогах;
- группировать файлы по каким-либо признакам в одном каталоге, по контексту связывая эти признаки с соответствующим названием;

- обеспечивать более удобный и быстрый, по сравнению с линейным списком файлов, доступ пользователя к файлам.

**2. Организация виртуальной структуры хранения файлов**

Порядок соподчинения каталогов и файлов в древовидной структуре определяется пользователем и соответствует практике иерархической классификации информации. Но при классификации один и тот же объект, одно и то же понятие включается, как правило, во множество иерархических структур, в то числе это относится и файлам и структурам файлов. В настоящее время для того, чтобы организовать новую структуру с файлом или множеством файлов, необходимо создать новую древовидную структуру и скопировать в неё нужное подмножество файлов. Такой способ имеет два существенных недостатка:

– нерационально используется память компьютера (в системе присутствует множество экземпляров одного и того же файла);

– появляется необходимость согласования множества экземпляров файлов при изменении одной из копий файла (актуализация файлов).

Рассмотрим пример. Пусть имеются электронные книги, характеризуемые названием, автором и годом издания. Мы можем их разместить по каталогам следующим образом – сначала разбить книги по годам издания, а потом по фамилиям авторов (рис. 2). Это соответствует классификации, по которой можно узнать в какие годы и какие авторы публиковались.

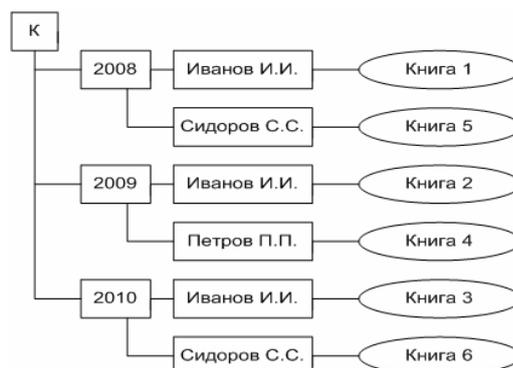

*Рис. 2 Пример систематизации по годам и авторам*

Второй способ размещения может соответствовать ситуации, когда нас будут интересовать конкретный автор и его публикации по годам (рис. 3).

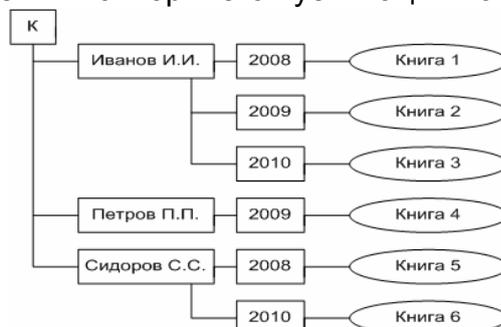

*Рис. 3 Пример систематизации по авторам и годам*

Предлагаемый в статье подход позволяет избежать этих недостатков при построении множества системообразующих структур. Рассмотрим этот подход более подробно. Сформируем линейный список файлов хранения. Каждому файлу, имеющему уникальное содержание, сопоставим уникальный номер, дублирующие экземпляры при этом должны быть исключены. Далее на основе списка файлов построим древовидную структуру, которая будет состоять из **виртуальных** каталогов и ссылок на файлы. Она внешне будет выглядеть так же, как и стандартная древовидная структура операционной системы. Однако то, что каталоги являются **виртуальными** и содержат только ссылки на уникальный номер файла, и доступ к файлам не привязан к длинному имени, позволяет построить множество деревьев на одном поле файлов. Для одного и того же набора файлов строятся различные деревья по различным признакам с учетом принятой систематизации. На рис. 4 линейный список представлен семью файлами, для которого построены два **виртуальных** дерева. Многоточие означает, что может быть построено произвольное количество таких деревьев. Буквами '**Ф**' и '**К**' на рисунке обозначены соответственно **Ф**айлы и **К**аталоги.

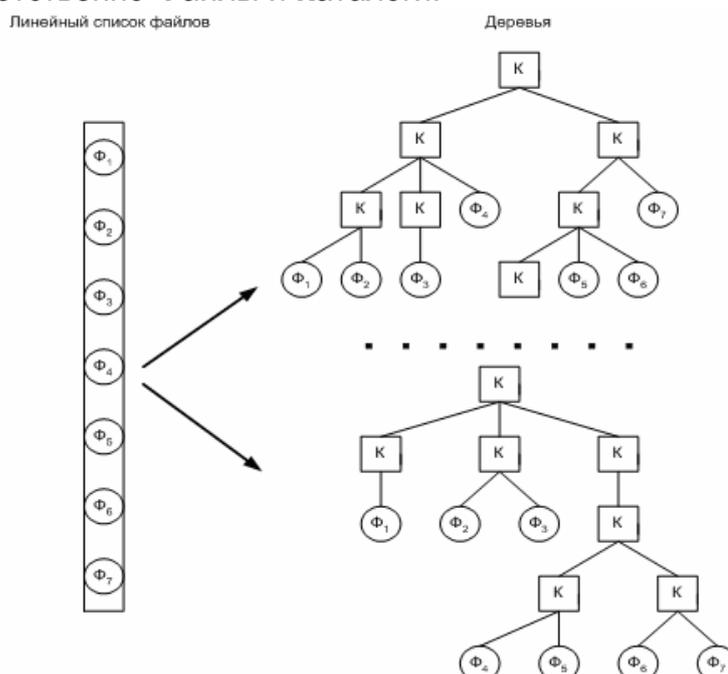

*Рис. 4 – Виртуальная структура файлов*

Непересекающиеся по классификации файлы и структуры можно выделить в отдельное множество, например «классики серебряного века». Такой набор файлов с уникальными номерами и виртуальными структурами будем называть **разделом.** Таким образом, при необходимости всё пространство файловой памяти может быть разбито на произвольное количество независимых по содержанию **разделов**.

Построение виртуальных структур в разделах может идти двумя способами:
– ручное создание каталогов и «привязка» к ним файлов;
– автоматическое создание каталогов по предварительно заданным признакам файлов, по которым производится систематизация.

При первом способе, каждое **виртуальное** дерево строится путём непосредственного связывания иерархии **виртуальных** каталогов. Каждый файл может разместиться в дереве только один раз. Построенное **виртуальное** дерево соответствует древовидной структуре обычной файловой системы, различие заключается в том, что файлы идентифицируются своими уникальными номерами, а не полным именем, определяющим путь к файлу.

Второй способ состоит в построении **виртуальных** деревьев по задаваемым заранее признакам файлов, которые задаются заранее (в нашем примере годы издания и фамилии авторов). Пусть имеется **раздел**, в котором находится определённое количество файлов. Все файлы имеют общее множество признаков. Признаки в базе данных имеют название признака, а самому признаку присваивается определённое значение. В нашем случае признаку названному как годы издания присваивается конкретный год, а признакам фамилия писателей - фамилия конкретного автора. Пусть общее количество различных признаков равно *a*. А общее количество файлов равно m. Тогда для *i*-ого файла значение признаков можно представить как строку $n_{i1}, n_{i2}, \ldots, n_{ia}$. Таким образом, все признаки всех файлов раздела можно представить матрицей *L*:

$$L = \begin{pmatrix} n_{11} & n_{12} & \ldots & n_{1a} \\ n_{21} & n_{22} & \ldots & n_{2a} \\ \ldots & \ldots & \ldots & \ldots \\ n_{m1} & n_{m2} & \ldots & n_{ma} \end{pmatrix},$$

*где $n_{ij}$ – значение j-ого признака для i-ого файла.*

Для построения какого-либо **виртуального** дерева в определённой последовательности выбирается перечень признаков, что и будет определять иерархию полученного дерева. На первом шаге дерево представляет собой один корневой каталог (рис. 5).

*Рис. 5 Построение дерева (первоначальный вид)*

Пусть выбраны *h* признаков и b файлов, что соответствует матрице *M*:

$$M = \begin{pmatrix} n_{11} & n_{12} & \ldots & n_{1h} \\ n_{21} & n_{22} & \ldots & n_{2h} \\ \ldots & \ldots & \ldots & \ldots \\ n_{b1} & n_{b2} & \ldots & n_{bh} \end{pmatrix},$$

*где $n_{ij}$ – значение j-ого признака для i-ого файла.*

Матрица *M* может быть получена из матрицы *L* удалением части столбцов и строк, а также возможной перестановкой столбцов выбранных признаков. Далее, в соответствии со значениями первого столбца матрицы *M* формируются элементы 1 яруса дерева, где имя каталога соответствует значению из этого списка. В этом списке каждое значение выбирается только один раз. Пусть число выбранных значений признаков равно *q*. Тогда к корневому каталогу добавятся *q* каталогов (рис. 6).

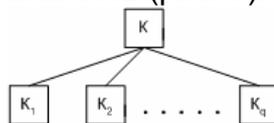

*Рис. 6 – Построение дерева (1-ый этап)*

Пуст на следующем шаге из значения признаков второго столбца матрицы *M* по такому же принципу формируются *r* подкаталогов для каталога К$_1$, *t* подкаталогов для каталога К$_2$, и наконец *e* подкаталогов для каталога К$_q$. Таким образом, имеем структуру дерева изображенного на рисунке 7.

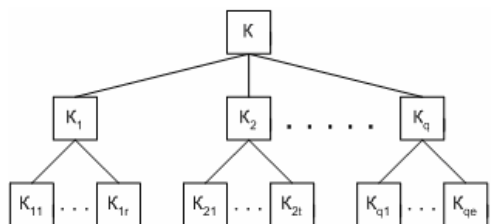

*Рис. 7 Построение дерева (2-ой этап)*

Далее этот процесс повторяется для последующих столбцов матрицы *M*. Глубина дерева соответствует количеству наименований признаков или просто признаков. Когда будут пройдены все *h* столбцов, получится иерархическое дерево. Количество уровней будет равно *h*+1. После этого к дереву прикрепляют файлы, которые соответствуют полному пути по построенному дереву или, что то же самое, значению всех признаков из базы данных, рис. 8. Весь процесс такого построения происходит автоматически, пользователь лишь определяет перечень требуемых признаков в необходимой последовательности. Для одних и тех же файлов может быть построено любое количество деревьев.

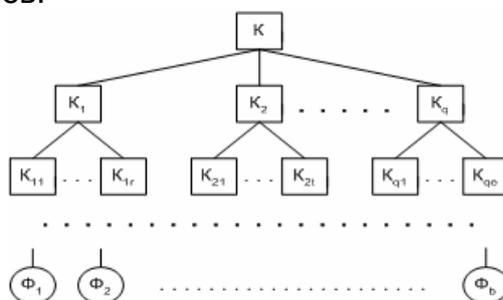

*Рис. 8 Построение дерева (последний этап)*

## Заключение

Таким образом, в статье предложен новый подход к систематизации и организации файловой структуры – **разделы**, **виртуальные** структуры. Он реализован в виде простого программного комплекса. Этот подход позволяет на основе единого набора файлов систематизировать записанную в них информацию, в виде множества различных файловых структур двумя способами. Очевидно, что такое построение файловой системы позволяет оптимизировать использование дискового пространства компьютера, а также избежать необходимости актуализировать копии файлов, что требуется при стандартном построении файловой структуры.